# Stay hydrated: Basolateral fluids shaping tissues

Markus Frederik Schliffka[1,2] and Jean-Léon Maître[1,*]

[1] Institut Curie, PSL Research University, Sorbonne Université, CNRS UMR3215, INSERM U934, Paris, France
[2] Carl Zeiss SAS, Marly-le-Roi, France
[*] Correspondence to: jean-leon.maitre@curie.fr

**Abstract**
During development, embryos perform a mesmerizing choreography, which is crucial for the correct shaping, positioning and function of all organs. The cellular properties powering animal morphogenesis have been the focus of much attention. On the other hand, much less consideration has been given to the invisible engine constituted by the intercellular fluid. Cells are immersed in fluid, of which the composition and physical properties have a considerable impact on development. In this review, we revisit recent studies from the perspective of the fluid, focusing on basolateral fluid compartments and taking the early mouse and zebrafish embryos as models. These examples illustrate how the hydration levels of tissues are spatio-temporally controlled and influence embryonic development.

**Fluid compartments**

During animal development, the embryo changes its shape following a carefully orchestrated program. This transformation results from combined cellular processes such as cell growth, proliferation, deformation, neighbor exchange and migration [1,2]. Often, these movements lead cells to form coherent compartmentalizing structures [3]. These compartments are bounded by a barrier, typically an epithelial layer, and can then gain further complexity [4]. Additionally, all living cells are immersed in fluid, of which the volume, composition and movement are controlled within larger compartments such as lumens, organs or vessels. This fluid is crucial for cell and tissue homeostasis, as its properties determines many cellular functions, from metabolism to signaling [5–11].

We define fluid compartments as units within which fluid freely permeates. Fluid compartments are bounded by a barrier, which controls the exchange of solutes and water (Fig. 1). Typically, this barrier is made of epithelial or endothelial cells, which are apico-basally polarized and form tight junctions [12,13]. Apico-basal polarity allows for directed transport, while tight junctions between the barrier cells ensure that fluid and solutes cannot freely diffuse over the compartment barrier [14]. Together, apico-basal polarity and tight junctions permit the definition of chemically and physically distinct fluids in the apical and basolateral compartment (Fig 1). Thereby, tissues can confine chemical signaling to the apical or basolateral compartments [9,15,16]. For example, in the zebrafish lateral line, cells restrict FGF signals to a few cells of the organ, which share a small apical lumen of a few microns in diameter [17]. In other cases, such as in flat gastruloids, Nodal signaling can be perceived differently by cells if applied within the apical or basolateral compartments as a result of polarized receptor localization [18]. Similarly, physical signals, such as the shear stress produced by contraction-generated flows, can be confined to a compartment. For example, shear stress controls heart valve differentiation during zebrafish heart development [19,20]. Fluid flow in other contexts, e.g. interstitial flow [21,22] or cilia-mediated flow also generate shear and may induce a cellular response if sufficiently strong [23]. Finally, osmotic and hydrostatic pressures are able to set distinct volumes to compartments and organs, such as the chick eye [24], zebrafish brain ventricle [25] or mouse blastocyst [26].

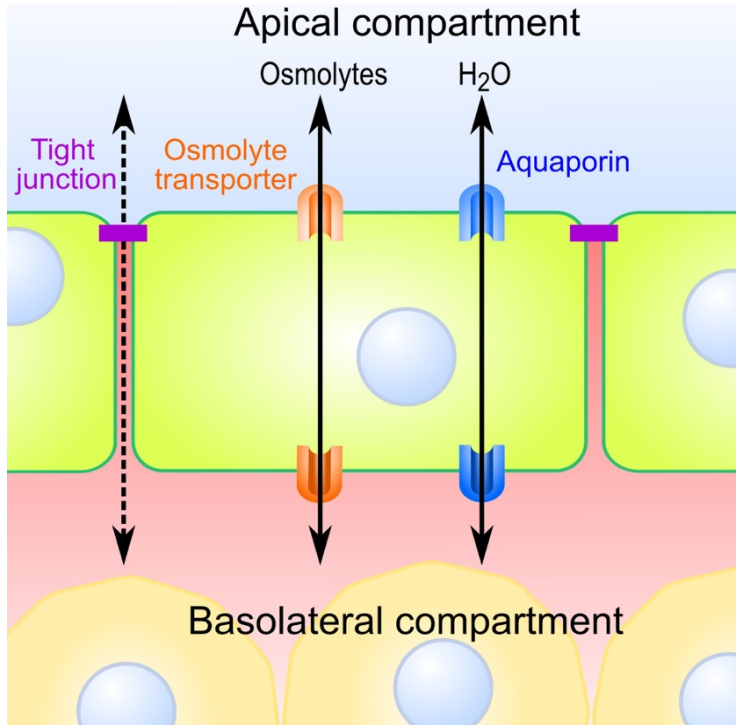

*Figure 1: Cellular control of fluid compartments.*
*Fluid movement between compartments is controlled by barriers such as epithelia or endothelia. Transmembrane pumps, transporters and channels control the distribution of osmolytes in both compartments [27]. Water follows the osmotic gradient established across the cellular barrier by flowing through aquaporins, which are water-specific transmembrane channels [28,29]. The sealing of the cell barrier is maintained by tight junctions, which control paracellular exchange [14].*

Previous reviews thoroughly covered crucial aspects related to intercellular fluid such as apical lumen formation [12,30,31] or the generation and sensing of fluid flow [32,33]. Here, we focus on recent findings on fluid permeating the basolateral compartment. Unlike apical compartments, basolateral compartments typically contain less fluid and are populated with cells which adhere to each other and to the extracellular matrix [34]. In the following, we will discuss recent findings on how basolateral fluid is controlled and how it impacts the morphogenesis of the mouse blastocyst and of the zebrafish gastrula.

**The mammalian blastocoel: a blueprint for basolateral lumens**

The blastocoel is the first lumen to form during the early development of several groups of animals such as echinoderms, amphibians or mammals [35–37] The blastocoel is a basolateral lumen, which may come across as unfamiliar when most studies on lumens focus on apical ones [12,30,31,38]. Yet, MDCK cells, a canonical model for apical lumen formation [31,39], can form "inverted cysts" with a lumen on their basolateral side if cultured in suspension [40,41] or form domes when in a monolayer by pumping fluid towards their basolateral interface [42]. Such epithelial structures with "inverted" polarity may turn out not to be that uncommon in physiological and pathological settings [43]. Apical and basolateral cysts will likely share many features, with one key difference: basolateral lumens form at the adhesive interface of cells. To study this aspect, the mouse blastocyst, forming during preimplantation development, can serve as an ideal experimental system.

The blastocyst consists of an epithelial envelope, the trophectoderm (TE), enclosing the inner cell mass (ICM) and the blastocoel [44,45]. The blastocoel is a basolateral fluid compartment, in which cells from the ICM typically gather into a coherent structure, but can also be found dispersed throughout the embryo against the TE envelope, e.g. in the elephant shrew [46]. Prior to blastocoel formation, the mouse embryo first compacts, pulling its cells into close contact and reducing intercellular fluid [47] (Fig. 2). Then, surface cells acquire apico-basal polarity and seal their basolateral compartment by expanding their apical domain and fastening their tight junctions [48,49]. This allows polarized transport of osmolytes and fluid towards the basolateral compartment [50–52] (Fig. 2).

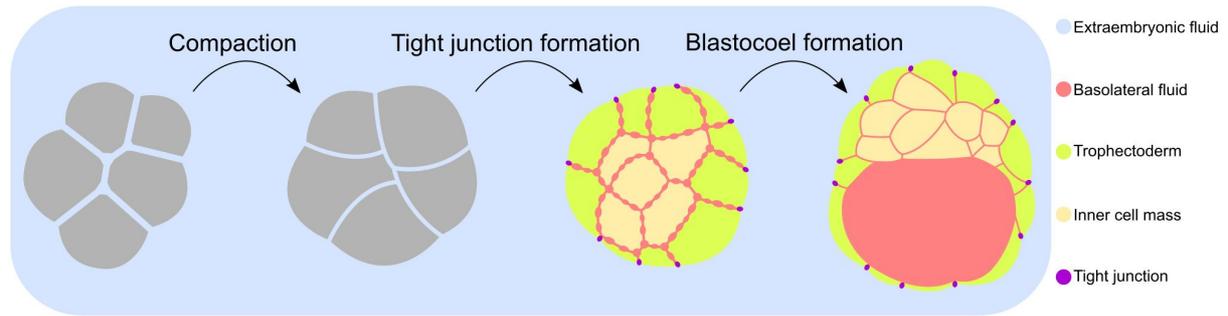

*Figure 2: Basolateral lumen formation in the mouse blastocyst. During compaction, the mouse embryo minimizes the amount of intercellular fluid between cells (blue). After the fourth cleavage, the embryo establishes tight junctions (purple) between surface cells as they become a polarized epithelium (green). This process separates the intercellular (red) and extraembryonic (blue) fluid compartments. Then, directed osmolyte and water transport leads to fluid accumulation in the basolateral compartment* [50–52]. *The pressurized fluid fractures cell-cell contacts into micron-size lumens. Tensile and adhesive mechanical properties of the tissue constrain the fluid into a single lumen, which becomes the blastocoel* [53]. *As the blastocoel expands, it mechanically challenges the integrity of the compartment by stretching the epithelial envelope and testing the tight junctions* [26,54].

At the onset of blastocoel formation, fluid accumulates in the intercellular space of the mouse embryo and cell-cell contacts become separated by hundreds of short-lived micron-size lumens [53]. Analogous structures were previously described when fluid is flushed on the basolateral side of MDCK monolayers [55]. The pressure of the fluid, in the range of hundreds of pascals, is able to fracture adherens junctions, resulting in short-lived fluid pockets between cell-cell contacts. In the blastocoel, the pressure is of similar magnitude, as measured on rabbit [50] and mouse embryos [26,53,56], suggesting that basolateral lumens forming at adhesive interfaces can nucleate by hydraulic fracturing of cell-cell contacts.

Micron-size lumens eventually disappear except for one that becomes the blastocoel [53]. In epithelial monolayers *in vitro*, cracks formed by hydraulic fracturing seal back within minutes through a mechanism that depends on actomyosin contractility [55]. Analogous fluid gaps can be seen in early xenopus embryos when contractility is affected [57]. In the mouse embryo, differences in contractility, such as the one existing between TE and ICM cells [58], can guide the fluid towards the most deformable tissue, the TE [53] (Fig 2). The ability of cell contractility to guide intercellular fluid appears as a general feature for apical [25,59,60] or basolateral compartments [53]. In addition to contractility, other mechanical properties of tissues can direct basolateral fluid. During hydraulic fracturing, when the fluid percolates the material, it will follow the path of least resistance. When adhesion molecules are patterned, the fluid will collect in the least adhesive part of the embryo [53]. Studies on the mechanical stability and molecular regulation of junctions under hydrostatic stresses will be needed to understand how basolateral fluids shape tissues. Additional mechanisms, such as localized fluid transport, could be at play and will require further studies. Optogenetic tools, such as light-gated ion channels [61], could constitute a prime instrument in this investigation.

Once the blastocoel is the last remaining lumen, it continues expanding as fluid continues accumulating. This expansion can be counteracted by leakage of the luminal fluid through defective tight junctions [62–64], leading to the collapse of the blastocyst [26,65]. The origin of this leakage is manifold. Cell divisions can challenge the integrity of the epithelium [26,56] as daughter cells need to assemble new adherens and tight junctions [66,67]. Similar to the hydraulic fracturing of adherens junctions, excessive pressure could also rupture the tight junctions of the TE [49,67]. In addition, the ability of TE cells to stretch could also be limiting, with unusual mechanical behaviors potentially at play to reduce this effect [42]. Once compromised, the tightness of the epithelium and the homeostasis of the fluid compartments need to be restored, which can be achieved by modulating contractility in fractured adherens and tight junctions [55,67]. After tight junction resealing, the epithelium can proceed its expansion until the next collapse [26,54]. Cycles of swelling and collapse have been proposed to determine the size of fluid-filled lumens and organs. This principle has been studied in models such as the regenerating hydra [68,69] and the blastocyst [26,54]. With blastocysts reaching about 200 $\mu$m in diameter for the mouse [70] and up to 3 mm for the rabbit [71], the mammalian embryo constitutes an ideal playground to explore how fluid compartments control organ size.

**The basolateral fluid compartment enabling zebrafish gastrulation**

During gastrulation, cells sculpt the embryo by establishing the three germ layers and the three axes of the body plan. This process essentially relies on cell-cell contact rearrangements within tissues with varying degrees of cell density and mechanical properties [72,73]. In zebrafish, this process takes place in a basolateral compartment, which to some extent corresponds to the blastocoel found in other groups of animals such as amphibians [74]. This compartment is bounded by the enveloping layer (EVL), a squamous epithelium which dynamically controls the composition and localization of intercellular fluid [75,76]. The amount of fluid in this basolateral compartment varies among teleost species, with high cell density in zebrafish and comparably much fewer cells within a voluminous basolateral compartment in killifish [77]. We now discuss three sequential morphogenetic events occurring within the same basolateral fluid compartment delimited by the same boundary: the EVL.

Zebrafish gastrulation begins with doming, a process during which cells intercalate radially to form a dome-like structure on top of the yolk syncytial layer (Fig. 3). Changes in the surface tension of the EVL powers their rearrangement [75,78]. Throughout doming, cell density is regulated in time and space. Importantly, cell divisions affect cell-cell contact stability differently at the animal pole and at the equator. Cells near the margin become more packed while those close to the animal pole of the embryo become less dense [79]. This is accompanied by a redistribution of the intercellular fluid, which becomes more prominent at the animal pole (Fig. 3). If the fluid compartment boundary, the EVL, does not function properly, fluid fails to accumulate at the animal pole and doming is impaired [75,76]. What controls the redistribution of the fluid throughout the compartment and how it affects cell intercalation remains unclear. The friction between cells and the lubricating action of the fluid may constitute interesting aspects to investigate these questions. After doming is complete, the three germ-layers form when mesoderm and endoderm cells break off inward and leave the ectodermal layer [72]. Differences in contractility among germ layer progenitors can drive their separation [80] by controlling cell-cell contact formation and stability [81]. These differences in cell contractility crucially depend on the composition of the intercellular fluid. Changing the osmolarity of the medium can impact cell mechanics [82,83], and osmolarity is a decisive parameter in the sorting of the zebrafish germ layers [6]. How the composition of the intercellular fluid evolves and how this precisely impacts gastrulation remains to be determined. The function of the EVL in controlling the basolateral fluid composition and volume is likely to be key.

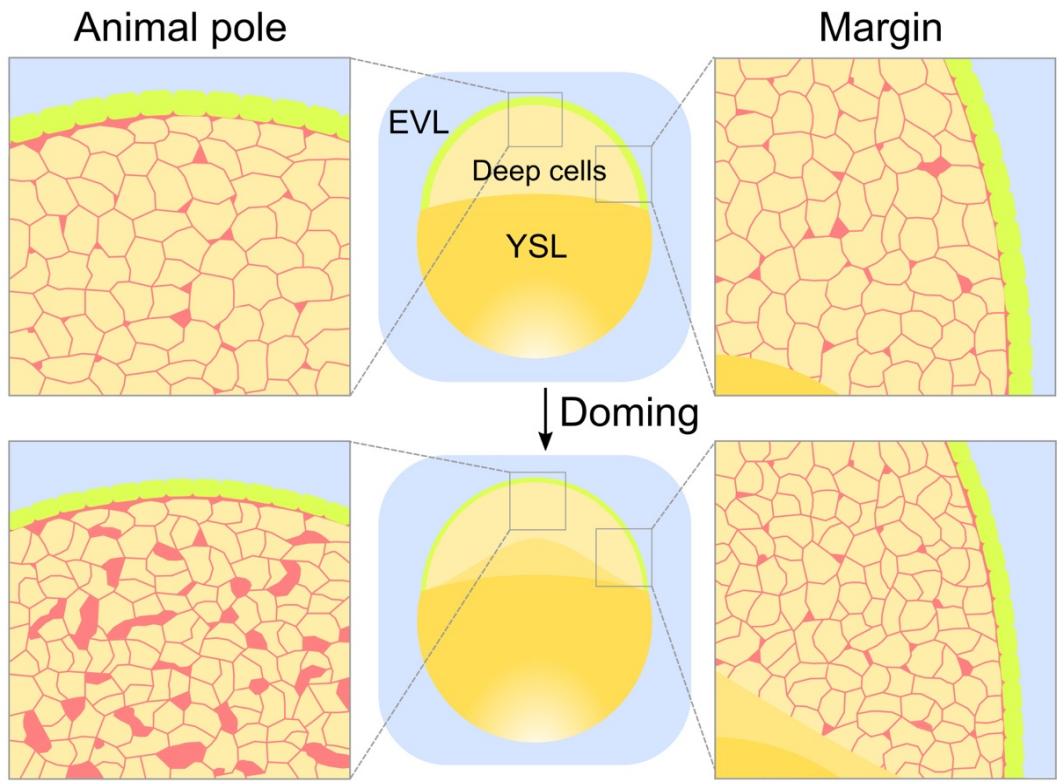

***Figure 3: Intercellular fluid distribution during doming in zebrafish embryos.*** *At the onset of zebrafish gastrulation, surface cells become epithelial, separating the intercellular from the extraembryonic fluid. As this surface epithelium increases surface tension, deep cells are forced to rearrange radially [75] and intercellular fluid follows. However, because cell-cell adhesion is stronger at the margin of the tissue (right) than at the animal pole (left), fluid accumulates at the animal pole and the marginal tissues becomes less hydrated [79]. EVL: enveloping layer. YSL: yolk syncytial layer.*

After germ layer separation, the embryo elongates its anteroposterior axis. Concomitantly, paraxial mesoderm cells from the elongating tailbud compact into somites [84,85] (Fig. 4). As mesoderm cells progressively incorporate into the presomitic mesoderm, they become less motile [86]. Although cell density appears homogeneous, this translates into higher tissue stress in regions with lower cell-cell rearrangements [87]. This phenomenon was described as a jamming transition during which the tissue becomes frozen due to cells packing themselves into somites [88]. Another way to describe this phenomenon is that when the tissue becomes too dry, the lack of lubrication prevents cells from rearranging and causes higher tissue stress. If cell-cell adhesion is defective, cells loosen up and the tissue becomes infiltrated with intercellular fluid, concomitantly with the lowering of the tissue stress [88]. How fluid is managed throughout the tissue during this process and how the hydration state of the tissue favors cell movement will need further investigation. Importantly, changes in the amount of fluid between cells in space and time is concomitant with changes in the mechanical property of the entire tissue [75,79,88]. The relative contribution of the fluid and of the cells to these properties remain to be determined.

**Concluding remarks**
In this review, we highlight the regulation and function of basolateral fluids during development. While previous research predominantly focused on forces generated directly by cells and tissues, comparably less attention has been given to the invisible fluid in which cells are immersed. We gather that basolateral compartments can be more or less hydrated in space and time (Fig. 4). For example, the tail of the zebrafish embryo becomes dryer as cells pack themselves anteriorly while the preimplantation embryo becomes more and more hydrated over time until it forms a lumen. These changes in the level of hydration of the tissue impact numerous properties, from the ability of cells to rearrange to the diffusion of morphogens. We hope the recent advances such as the ones brought by studying the early mouse and fish embryos will motivate new research in this direction. This will require developing new tools to visualize the fluid properties and manipulate them, as well as new frameworks to precisely delineate the mechanisms by which tissue hydration controls development.

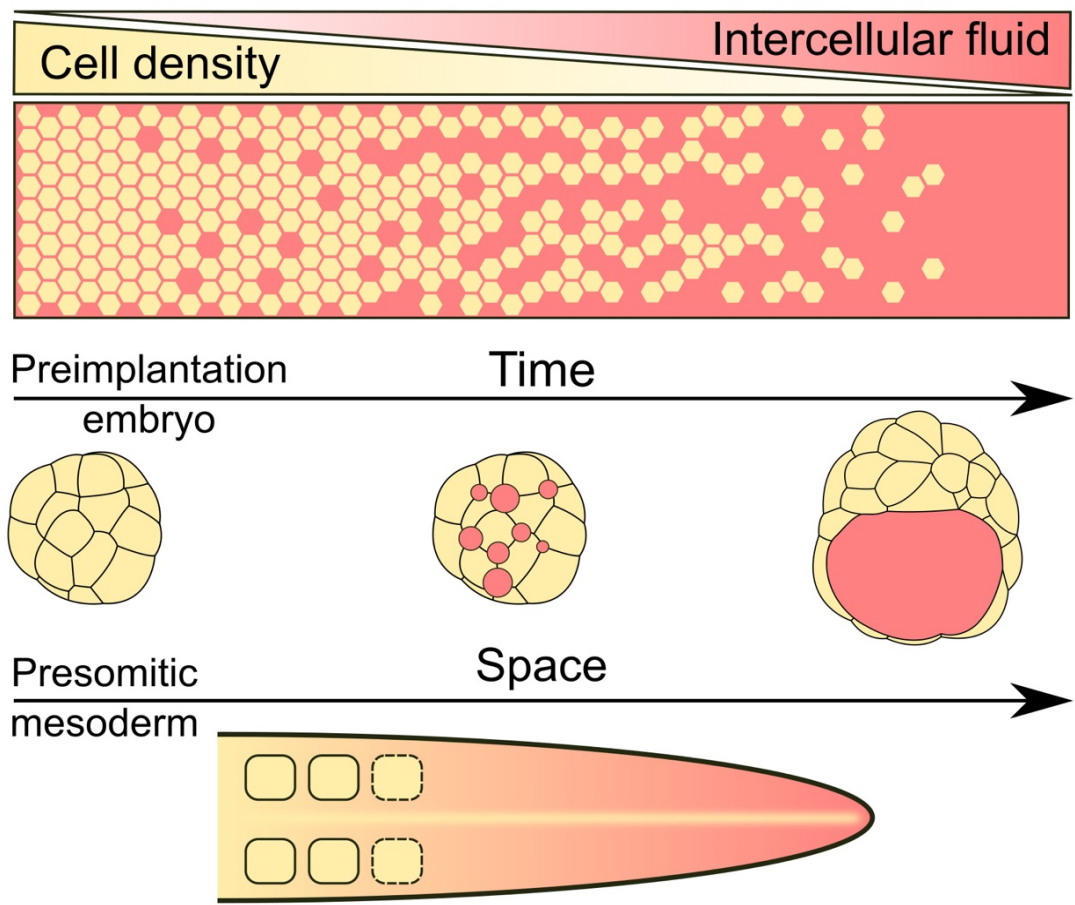

*Figure 4: Spatial and temporal control of tissue hydration during morphogenesis.*
*The level of hydration of tissues changes in space and time during development. The inner cells of the mouse embryo are initially densely packed. Fluid accumulates inside the embryo and separates the inner cells until it begins to collect in one single pocket, which becomes the blastocoel* [53]. *The blastocoel is now considered a lumen, a purely fluid-filled compartment. Dense tissues such as anterior paraxial mesoderm contain less fluid between their cells than posterior presomitic mesoderm. Cell movement is facilitated in the hydrated posterior and jammed in the anterior* [88].

**Acknowledgments:**
We thank Julien Dumortier, Diane Pelzer and Ines Cristo for fruitful discussions and comments on the draft of the review. Research in the lab of JLM is funded by the ATIP-Avenir program, an ERC-2017-StG 757557 and Labex DEEP (ANR-11-LBX-0044). MS is currently employed by Carl Zeiss SAS.